\documentclass[prl,twocolumn,aps,tightenlines,showpacs,floatfix]{revtex4}

\usepackage{graphicx}

\usepackage{dcolumn}     

\begin{document}
 
\title{Quantum Monte Carlo calculations of neutron-alpha scattering}

\author{Kenneth M. Nollett}
\email{nollett@anl.gov} 
\author{Steven C. Pieper}
\email{spieper@anl.gov}
\author{R. B. Wiringa\vspace*{-.15in}}
\email{wiringa@anl.gov}
\affiliation{Physics Division, Argonne National Laboratory, 
Argonne, Illinois 60439\vspace*{-.10in}}

\author{J. Carlson}
\email{carlson@lanl.gov}
\author{G. M. Hale\vspace*{-.15in}}
\email{ghale@lanl.gov}
\affiliation{Theoretical Division, Los Alamos National Laboratory,
Los Alamos, NM 87545}

\date{\today}
 
\begin{abstract}
We describe a new method to treat low-energy scattering problems in
few-nucleon systems, and we apply it to the five-body case of
neutron-alpha scattering.  The method allows precise calculations of
low-lying resonances and their widths.  We find that a good
three-nucleon interaction is crucial to obtain an accurate description
of neutron-alpha scattering.
\end{abstract}


\maketitle

There has been significant progress recently in understanding the
ground and low-lying excited states of light nuclei through
microscopic calculations with realistic two- and three-nucleon
interactions~\cite{WPCP00,PVW02,ncsm}.  These studies have highlighted
the importance of including a three-nucleon interaction to obtain
correct overall binding of the p-shell nuclei~\cite{WPCP00}, the
ordering of states in $^{10,11,12}$B~\cite{ncsm}, and the stability of
neutron-rich nuclei~\cite{PPWC01}.  Modern calculations have been very
successful in reproducing a large number of the
experimentally-observed nuclear levels up to mass 12.

A wealth of additional experimental information is available in the
form of low-energy scattering and reaction data.  Very narrow
low-energy resonances have been treated in the calculations discussed
above as essentially bound states, but non-resonant scattering and
broad resonances have not been adequately addressed.  These cases
require the development of methods to compute 
scattering states, which in
turn will form the foundation for a quantitative, microscopic theory
of low-energy nuclear reactions on light nuclei.  
Such a theory would be
useful to astrophysics, because experimental determinations of crucial
reaction rates are often difficult and sometimes impossible.

Here we review the method previously used for quantum Monte Carlo
(QMC) calculations of low-energy reaction properties, present an
improved method, and use it to study low-energy neutron-alpha
scattering as a five-body problem.  We evaluate the two low-lying
$p$-wave resonances with $J^\pi$=$3/2^-$ and $1/2^-$, respectively, as
well as low-energy $s$-wave $(1/2^+)$ scattering.

In each case, we calculate the phase shift as a function of energy for
the Argonne $v_{18}$ (AV18) two-nucleon potential~\cite{WSS95} alone,
and with either the Urbana IX (UIX) model~\cite{PPCW95} or Illinois-2
(IL2) model~\cite{PPWC01} three-nucleon potential added.  We find
significant differences in the calculated phase shifts among these
Hamiltonians.  The AV18 and AV18+UIX models produce too little
splitting of the $3/2^-$ and $1/2^-$ states, while AV18+IL2 reproduces
their energies and widths very well.  All three models provide good
matches to the low-energy $s$-wave cross sections.  The results
demonstrate that experimental phase shifts can be reproduced using
realistic interactions with simple modifications to the computational
method used for bound states.  They also suggest that scattering
calculations could provide sensitive constraints on models of the
three-nucleon interaction.

In general it is very difficult to treat quantum scattering problems
with more than a very few constituents.  Often there are many initial
or final states, and it is not yet possible to discretize the
Schr\"odinger equation and solve directly for the scattering
states. Correct treatment of the boundary conditions for more than two
outgoing constituents is also complicated.

For low-energy scattering, though, there are often only a few
two-cluster channels open.  For one open channel of total angular
momentum $J$ and orbital angular momentum $L$, the wave function at
large separation $r$ of clusters with internal wave functions
$\Phi_{c1}$ and $\Phi_{c2}$ will behave as
\begin{equation}
\label{eqn:asymptotic}
    \Psi  
\propto
     \frac{1}{kr} 
\left\{\Phi_{c1}\Phi_{c2} Y_{L}\right\}_J
\left[\cos \delta_{JL} F_L(kr) 
+ \sin\delta_{JL} G_L(kr)\right] \ ,
\end{equation}
where $\delta_{JL}$ is the phase shift, $ k = \sqrt { 2 \mu E /
\hbar^2 }$, $E$ is the scattering energy in the c.m. frame, $\mu$ is
the reduced mass, $F_L$ and $G_L$ are respectively regular and
irregular real solutions of the Schr\"odinger equation with zero
nuclear potential, $Y_L$ are spherical harmonics, and $\{\cdots\}_J$
denotes angular momentum coupling.

Previous quantum Monte Carlo calculations of scattering states
converted the scattering problem to an eigenvalue problem by imposing
a boundary condition $\Psi(r\!=\!R_0) = 0$ at a surface radius
$R_0$ beyond the range of the nuclear interaction, and then solving
for the energy within this nodal surface~\cite{CPW84,AK84}.  
Eq.~(\ref{eqn:asymptotic}) then gives 
\mbox{$\tan \delta_{JL} = - F_L( kR_0 )/G_L( k R_0 )$}.
Finding $E$ as a function of $R_0$ is then equivalent to determining
$\delta_{JL} (E)$.

The zero boundary condition requires different $R_0$ for
different scattering energies.  Energies near threshold require very
large $R_0$, which can cause numerical difficulties.  For these
reasons, it is preferable instead to impose a logarithmic derivative
$\gamma$ on the wave function along the direction $\mathbf{\hat{n}}$
normal to the $r=R_0$ surface:
\begin{equation}
\label{eqn:logbc}
\mathbf{\hat{n}}\cdot \nabla_{\bf r} \Psi =
\gamma \Psi\ , ~~~\mathrm{at}\ r=R_0.
\end{equation}
Equation (\ref{eqn:asymptotic}) then gives $\delta_{JL}$ from
$E$, $R_0$, and $\gamma$.

We compute nuclear energies by using two QMC methods as successive
approximations.  The variational Monte Carlo (VMC) method uses a trial
wave function $\Psi_T(J^\pi;T)$ specified by variational parameters
that are adjusted to minimize the energy expectation value; Monte
Carlo techniques are used to evaluate the integrals over particle
position.  The method and the form of $\Psi_T$ are described in detail
in Refs.~\cite{WPCP00,PPCPW97}.  VMC calculations of $n$-$\alpha$
scattering were first made in the 1980s~\cite{CSK87}, by using a
$\Psi_T$ that explicitly went to zero at $R_0$.
In the present work, correlations inside $\Psi_T$ are constrained so
that it goes to the form Eq.~(\ref{eqn:asymptotic}) at large $r$ and
satisfies Eq.~(\ref{eqn:logbc}).

Green's function Monte Carlo (GFMC) takes the VMC trial state and evolves
it in imaginary time $\tau$,
\begin{equation}
\label{eqn:gfmc}
    \Psi ( \tau ) = \exp [ - (H-E_0) \tau ] \Psi_T \ ,
\end{equation}
so that $\Psi ( \tau \rightarrow \infty ) \propto \Psi_0$, the lowest
state with the specified quantum numbers.  The propagation time is
divided into many small steps of length $\Delta\tau$, and
Eq.~(\ref{eqn:gfmc}) is computed by iterating the small-time-step
Green's function,
\begin{equation}
G ({\bf R}^\prime, {\bf R}) = \langle {\bf
R}^\prime | \exp [ - (H - E_0) \Delta \tau ] | {\bf R} \rangle\ ,
\end{equation}
acting on samples of $\Psi_T$.  Here ${\bf R}=({\bf r}_1, {\bf r}_2,
\ldots, {\bf r}_A)$ is the spatial configuration of the nucleons, and
we suppress spin-isospin labels for simplicity.  
The nuclear GFMC method is described in Refs.~\cite{WPCP00,PPCPW97}.
GFMC calculations with the nodal condition at $R_0$ have been performed for
$n$-$\alpha$ scattering previously~\cite{C91} and also recently for
some atomic and condensed-matter problems~\cite{SC01,CMM02}.

In this work, we restrict the GFMC simulation to cluster separations
less than $R_0$ by rejecting proposed Monte Carlo steps that go
outside the boundary.  To enforce the boundary condition in
Eq.~(\ref{eqn:logbc}), we regard the wave function in the restricted
volume as the central region of a scattering wave function that fills
all space, with some energy specified implicitly by the boundary
condition.  Then the contributions to $\Psi(\tau)$ at the $(n+1)$th
step in $\tau$ come from inside and outside the $r<R_0$ region:
\begin{eqnarray}
&&\Psi_{n+1} ({\bf R}^\prime) = \int_{|{\bf r}| <R_0} 
                               d{\bf R}_{c1}\, d{\bf R}_{c2}\, d{\bf r}
           \ G ({\bf R}^\prime, {\bf R}) \Psi_n ({\bf R})  \nonumber \\
                            && +  \int_{|{\bf r}_e| > R_0} 
                               d{\bf R}_{c1}\, d{\bf R}_{c2}\, d{\bf r}_e
           \ G ({\bf R}^\prime, {\bf R}) \Psi_n ({\bf R}) \ .
\label{eqn:integral1}
\end{eqnarray}
Here ${\bf R}$ is written in terms of both internal cluster
coordinates $\mathbf{R}_{c1}$ and $\mathbf{R}_{c2}$, and
the cluster-cluster separations ${\bf r}$ and ${\bf r}_e$, which
respectively are contained inside, or extend outside, of the boundary
$R_0$.

The contribution of the outer region can be mapped to an integral over
the interior region by a change of variables,
${\bf r} = (R_0/r_e)^2 {\bf r}_e$ with $r_e = |{\bf r}_e|$,
in the second term of Eq.~(\ref{eqn:integral1}).
A wave function sample
at the point $\mathbf{R}^\prime$ is then the sum of one contribution
from the previous point $\mathbf{R}$ and one from an ``image point''
at $\mathbf{R}_e$, located outside the $r=R_0$ boundary:
\begin{eqnarray}
&&\Psi_{n+1} ({\bf R}^\prime)  =  \int_{|{\bf r}| < R_0}
                      d{\bf R}_{c1}\, d{\bf R}_{c2}\,   d{\bf r} 
           \ G ({\bf R}^\prime, {\bf R})  \nonumber \\
&& \times \left[ \Psi_n ({\bf R}) + 
     \frac{G ({\bf R}^\prime, {\bf R}_e)} {G ({\bf R}^\prime,{\bf R})} 
  \  \left( \frac{{ r}_e}{{ r}} \right)^3
  \  \Psi_n ({\bf R}_e) \right] \ .
\end{eqnarray}
Because $G({\bf R}^\prime,{\bf R})$ has short range, we
can use the boundary condition to obtain $\Psi_n ({\bf R}_e)$ by
linear extrapolation,
\begin{equation}
  \Psi_n(\mathbf{R}_e) \approx 
\left[1+\gamma \left(\mathbf{R}_e-\mathbf{R}\right)\cdot\mathbf{\hat{n}}\right]
\Psi_n(\mathbf{R})\ ,
\end{equation}
or by 
fixing the energy and extrapolating with
Eq.~(\ref{eqn:asymptotic}).  The GFMC simulation considers each
partition of the nucleons into clusters with separation ${\bf r}$.  If
the resulting $|{\bf r}-{\bf r}_e| < 1$~fm, then the image
contribution is used; otherwise the short range of $G$ makes it
insignificant.

This method has several advantages over the nodal boundary condition.
The most important is that $R_0$ can be set to a constant value,
independent of the scattering energy.  It is also possible to
calculate correlated energy differences in the GFMC method for
different values of the boundary condition if they are all computed
for the same $R_0$.  In such cases the path integrals differ only at
the surface, and the energy differences are more accurate than the
individual energies.

\begin{figure}[]
\includegraphics[width=3.25in,angle=0]{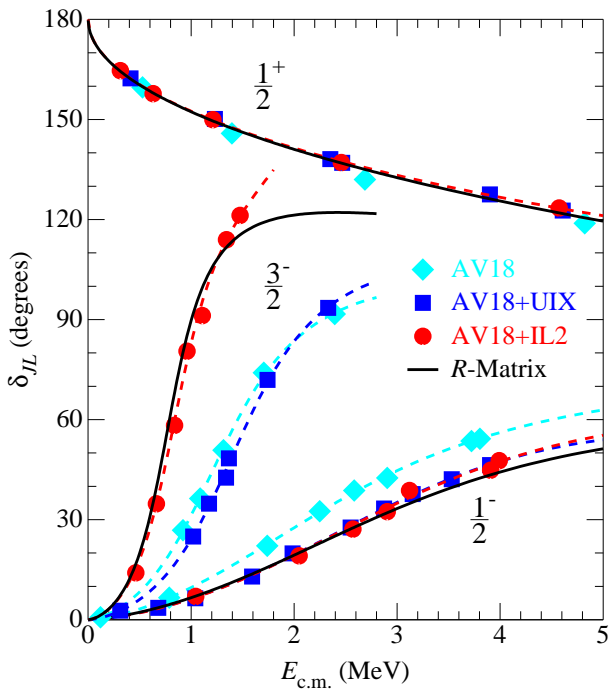}
\caption{(Color online) Phase shifts for $n$-$\alpha$ scattering.
Filled symbols (with statistical errors smaller
than the symbols) are GFMC results; dashed curves are
fits described in the text; and solid curves are 
from an $R$-matrix fit to data \protect\cite{hale}.}
\label{fig:shifts}
\end{figure}

Neutron-alpha scattering provides a convenient test of several
important properties of low-energy nuclear scattering.  There is no
bound state in the $A$=5 system, but the $3/2^-$ channel has a sharp
low-energy resonance.  The $1/2^-$ state is broader and higher in
energy, and the combination of these two states provides a simple
measure of spin-orbit splitting.  Since the alpha particle is so
tightly bound, simple single-channel scattering continues up to fairly
high energies.

The quantity most naturally given by a GFMC calculation is the total
energy of a system, and a precision in the vicinity of 1\% has been
the goal of our past calculations.  However, the quantity used to
compute phase shifts is the energy relative to threshold, so a
precision of 100 keV in this quantity is $\sim$0.3\% of the total
energy in the $^5$He problem.  To attain this, we must pay close
attention to the choice of $R_0$, construct a $\Psi_T$ as close as
possible to the desired GFMC solution, and use a less stringent GFMC
path constraint.

First, the only {\it a priori} constraint on $R_0$ is that it should
be ``beyond the range'' of the nuclear interaction. We find that our
GFMC result depends on $R_0$ at the level of about 100 keV (out of a
total of $\sim -28$ MeV) in going from $R_0=7$ fm to 9 fm.  We find no
further change going from 9 fm to 10 fm; the highest attainable
energy (corresponding to the state with a node at the surface) decreases
as $R_0$ increases, so we choose $R_0=9$ fm.  Future
calculations extending to energies beyond this maximum-energy state
should be analogous to previous calculations of multiple bound states
with the same quantum numbers~\cite{PWC04}.

Second, the GFMC energy also depends somewhat on the input $\Psi_T$.
We find it important to adjust pair correlations between particles in 
different clusters (between the $n$ and constituents of the $\alpha$ 
in this case) so that Eq.~(\ref{eqn:asymptotic}) is enforced at large 
cluster separation~\cite{NWS01}.  
We also adjust a parameter in $\Psi_T$ that corresponds to the scattering 
energy until it matches the final GFMC energy; this typically takes one or 
two iterations of the VMC and GFMC calculations to obtain a self-consistent 
result.

Finally, in all of our $A>4$ GFMC calculations, we use a path
constraint~\cite{WPCP00} on the GFMC walk to mitigate the Fermion sign
problem; we compute energy samples only after releasing the constraint
for some number of steps to avoid biasing the results.  We find that
stable results in our scattering calculations require the use of 80
unconstrained steps rather than the usual 20 to 40.

\begin{figure}[tb!]
\includegraphics[width=3.25in,angle=0]{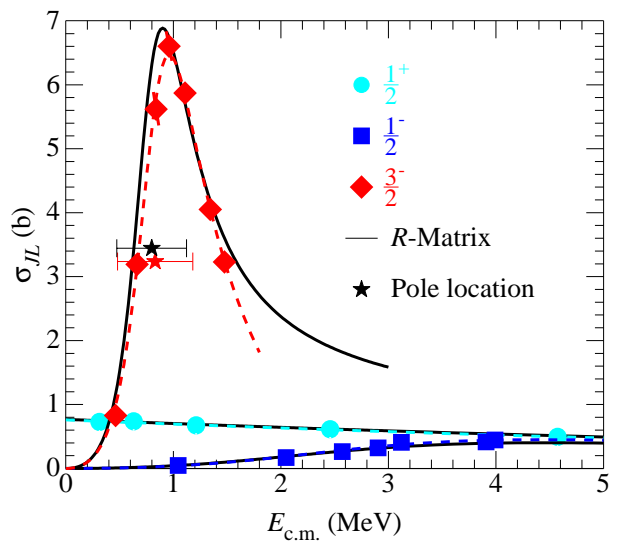}
\caption{(Color online) Calculated and $R$-matrix partial-wave cross
sections.  The calculations, shown with their Monte Carlo error bars,
are for the AV18+IL2 Hamiltonian.  Stars show the pole energies in
$3/2^-$ scattering for the $R$-matrix fit and for AV18+IL2, with the
bars indicating the imaginary part.}
\label{fig:crsec}
\end{figure}

In Fig.~\ref{fig:shifts} we present phase shifts for all channels,
computed with three different interaction models.  In each case the
AV18 potential is used as the two-nucleon interaction; in the second
(third) case the UIX (IL2) three-nucleon potential is added.  We also
show partial-wave total cross sections for the AV18+IL2 case in
Fig.~\ref{fig:crsec}.  We compare these results with those from a
multi-channel $R$-matrix analysis of the $^5$He system~\cite{hale}
that characterizes the measured scattering data very well
($\chi^2/$d.o.f. is 1.6).  Some of the resonance parameters from that
analysis are given in Refs.~\cite{CH} and \cite{a5eval}.  Because
there are more than 2600 data points in the analysis, the
uncertainties in the $R$-matrix phase shifts are likely to be much
smaller than the errors in the GFMC calculations.

We have made rational polynomial fits to \mbox{$\tan\delta_{JL}/k^{2L+1}$},
converted these to rational polynomials for the $S$-matrix, and used
these to find the poles of $S$.  These fits are shown as dashed curves
in the figures.  For each of the two $p$-wave states, we find just one
pole that is stable as the degrees of the polynomials are changed; we
identify these as the resonance poles.  For $3/2^-$ the poles are at
$1.19-0.77i$, $1.39-0.75i$, and $0.83-0.35i$ for AV18 alone, AV18+UIX,
and AV18+IL2, respectively, compared with $0.798-0.324i$~MeV from
analysis of the data~\cite{a5eval}.  The corresponding $1/2^-$ values
are $1.7-2.2i$, $2.4-2.5i$, and $2.3-2.6i$ MeV, compared with
$2.07-2.79i$~MeV.  The $1/2^+$ fits yield no stable pole, in agreement
with the lack of a resonance in this channel and with the $R$-matrix
analysis. We have quoted all pole locations, and the scattering length
below, so that there is an error of not more than 3 in the last
decimal place.

It is well known that realistic two-nucleon interactions alone provide
insufficient spin-orbit splitting in light nuclei~\cite{PP93,FM1}.
The figures and pole positions above confirm this: although there is
some spin-orbit splitting with AV18, it is less than half of what is
needed to explain the data.  The UIX three-nucleon potential includes
a two-pion-exchange term and a phenomenological short-range repulsion,
fitted simultaneously to the triton binding energy and the saturation
density of nuclear matter.  Adding UIX to AV18 increases the
spin-orbit splitting, but not enough to match the data.
  
In addition to the terms in the UIX model, the IL2 potential includes
a three-pion-ring exchange term and an additional two-pion-range term.
The strengths of all four terms were adjusted to fit seventeen energy
levels in nuclei up to $A$=8~\cite{PPWC01}.  Adding IL2 to AV18
induces even greater spin-orbit splitting in the fitted nuclei; the
figures and pole positions show that the experimental $^5$He splitting
is almost exactly reproduced by AV18+IL2.  For comparison, excitation
energies for this interaction have an RMS deviation of 600 keV from
experiment for the states to which it was fitted.  The widths of the
resonances are also well reproduced by AV18+IL2.

All three potentials produced essentially identical results for the $1/2^+$
case, in good agreement with the laboratory data.  The
$s$-wave scattering is dominated by a node in the $n$-$\alpha$
correlation at a position that roughly fixes the zero-energy
scattering length, and the wave functions for all potentials contain
similar structure.  All three are consistent with a scattering length
of 2.4 fm, compared with an experimental  scattering length
of 2.46 fm~\cite{a5eval}.

In summary, we have introduced a new QMC technique to calculate
low-energy scattering and applied it to neutron-alpha scattering.
Models of the three-nucleon interaction that have been successful in
describing bound levels of light nuclei also provide good descriptions
of these scattering states.  In the case of the most successful
potential, AV18+IL2, the description is particularly good.  The
sensitivity of the results in Fig.~\ref{fig:shifts} to the Hamiltonian
suggests that scattering calculations can provide important additional
constraints on the three-nucleon interaction.
We note that the energies obtained here for the $3/2^-$ pole are close
to those obtained by treating it as a bound state~\cite{PPWC01}, while
the pseudo-bound-state calculations of the broad $1/2^-$ state proved
less reliable.

Many extensions and applications of the computational method presented
here are possible.  Because these calculations are done in coordinate
space, including Coulomb interactions between the clusters will pose
no problems; the $F_L$ and $G_L$ of Eq.~(\ref{eqn:asymptotic}) are
then simply Coulomb, rather than Bessel, functions.  Extensions to
nucleus-nucleus (rather than nucleon-nucleus) scattering should also
be feasible.  Besides pure scattering problems and the use of
scattering wave functions to compute electroweak cross sections~\cite{mnsw06} 
and hadronic parity violation, 
the scattering method should be applicable to cases
with more than one open two-cluster channel.  For $N_\mathrm{ch}$ such
channels, a set of $N_\mathrm{ch}$ linearly independent solutions of
the Schr\"{o}dinger equation at the same energy determines the
$S$-matrix.  The solutions could be obtained either by varying the
boundary conditions at the surface, or by calculating the derivative
of the energy with respect to changes in the logarithmic derivatives.

\acknowledgments

We gratefully acknowledge valuable discussions with V. R. Pandharipande.  The
many-body calculations were performed on the parallel computers of the
Laboratory Computing Resource Center, Argonne National Laboratory and
the National Energy Research Scientific Computing Center.  The work of
KMN, SCP, and RBW is supported by the U. S. Department of Energy,
Office of Nuclear Physics, under contract No. DE-AC02-06CH11357.  The
work of JC and GMH is supported by the U. S. Department of Energy
under contract No. DE-AC52-06NA25396.


\end{document}